\documentclass[10pt, conference]{IEEEtran}
\normalsize

\usepackage[letterpaper, left=1.01in, right=1.01in, bottom=1in, top=0.75in]{geometry}

\usepackage[utf8]{inputenc}

\normalsize
\DeclareMathAlphabet\mathbfcal{OMS}{cmsy}{b}{n}

\usepackage[english]{babel}
\usepackage{amsmath,amssymb}
\usepackage{cite}
\usepackage{graphicx}
\usepackage{dsfont}
\usepackage{amsthm}
\usepackage{amssymb}
\usepackage{mdframed}
\usepackage{subcaption}
\usepackage{enumitem}

\newcommand{\idle}{\mathrm{i}}
\newcommand{\retx}{\mathrm{x}}
\newcommand{\new}{\mathrm{n}}
\newcommand{\sense}{\mathrm{n}}
\makeatletter
\def\blfootnote{\xdef\@thefnmark{}\@footnotetext}
\makeatother

\usepackage{lscape}
\usepackage{bbm}
\DeclareMathOperator*{\argmin}{arg\,min}

\newcommand{\Exp}[1]{\mathbb{E}\left[ #1 \right]} 
\newcommand{\Exppi}[1]{\mathbb{E}\left[ #1 \right]} 
\title{Reinforcement Learning to Minimize Age of Information with an Energy Harvesting Sensor with HARQ and Sensing Cost}

\author{
\IEEEauthorblockN{Elif Tu\u{g}\c{c}e Ceran, Deniz G{\"u}nd{\"u}z, and Andr\'as Gy\"orgy}
\IEEEauthorblockA{Department of Electrical and Electronic Engineering, Imperial College London\\
Email: \{e.ceran14, d.gunduz, a.gyorgy\}@imperial.ac.uk
}}

\begin{document}

\maketitle

\begin{abstract}
The time average expected age of information (AoI) is studied for status updates  sent from an energy-harvesting transmitter with a finite-capacity battery.  The optimal scheduling policy is first studied under different feedback mechanisms when the channel and energy harvesting statistics are known. For the case of unknown environments, an average-cost reinforcement learning algorithm is proposed that learns the system parameters and the status update policy in real time. The effectiveness of the proposed methods is verified through numerical results. 
\end{abstract}


\section{Introduction}

There has been a growing interest in minimizing the age of information (AoI) of energy harvesting (EH) communication systems \cite{Tan2015,Yates2015,Tan2017,Arafa2018,Arafa2017,Wu2018,Tan2018,Feng2018,ArafaPoor2018}. The AoI quantifies the staleness of the information at the receiver, and is defined as the time elapsed since the generation time of the most recent status update packet successfully received at the receiver.

Prior works have investigated online \cite{Tan2015,Tan2017,Tan2018} and offline \cite{Tan2015,Arafa2017} methods for different scenarios in order to optimize the timeliness of information under the energy causality constraints in EH systems. It is shown in \cite{Tan2017,Tan2018,ArafaPoor2018}  that the optimal policy is of a threshold type for a finite-size battery when the cost of sensing (monitoring) the status of a process is not considered or assumed to be zero. Until recently, prior literature in the AoI framework assumed that the cost of sensing (monitoring) the status of a process is negligible compared to the cost of transmitting the status update. However, in most practical sensing systems acquiring a new sample of the underlying process of interest also has a considerable energy cost. The sampling/sensing cost has been taken into account in \cite{Gong2018}, where a status update system with ARQ and an unlimited energy source is considered. Closed form expressions are presented for the energy consumption and AoI, assuming that a packet is re-transmitted until either it is received, or a prescribed maximum number of transmissions is reached.



In this paper, similarly to \cite{Gong2018}, we study a status update system considering both the sensing and transmission energy costs. We consider an EH transmitter, which uses the  energy harvested from the environment to power the sensing and communication operations. Moreover, we consider a hybrid automatic repeat request (HARQ) protocol, where the partial information obtained from previous unsuccessful transmission attempts is combined to increase the decoding probability.  

In our previous work, we studied status-update systems with HARQ under a transmission-rate constraint \cite{wcnc_paper,rl_paper,journal}. Here we consider the intermittent availability of energy and find the online status updating policy to minimize the long-term average AoI at the receiver, subject to the energy causality constraints at the transmitter. However, in many practical scenarios  the statistical information about either the energy arrival process or the channel conditions are not available or may change over
time \cite{Gunduz2014}. Previous work on EH communication systems without a-priori information on random processes governing the system exploited reinforcement learning (RL) methods in order to maximize throughput or minimize delay \cite{Blasco2013,Ortiz2016}.

To adapt the status-update scheme to the unknown energy arrival process and channel statistics, we propose a learning theoretic approach using RL algorithms. In particular, we consider a value-based RL algorithm, \emph{GR-learning} \cite{Gosavi2004}, and a policy-based RL algorithm, \emph{finite-difference policy gradient} \cite{policy_gradient}, and compare their performances with the relative value iteration (RVI) algorithm which assumes  a-priori knowledge on the system characteristics. We propose a suboptimal threshold policy and demonstrate that policy gradient algorithm exploiting the structural characteristics of a threshold policy outperforms GR-learning algorithm.  We investigate the effects of the EH process on the average AoI, and we show by simulations that temporal correlations in EH increase the average AoI significantly.  We compare the average AoI with EH with the average AoI under an average transmission constraint \cite{wcnc_paper} and demonstrate that the performance of RH transmitter approximates to the one with average transmission constraint for a battery with unlimited capacity and zero sampling/sensing cost.


\section{System Model}
\begin{figure}
\centering
\includegraphics[scale=0.3]{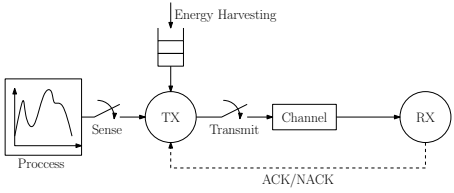}
\caption{An EH status update system over an error-prone link in the presence of ACK/NACK feedback.}
\label{fig:system}
\end{figure}

We consider a time-slotted status update system over an error-prone wireless communication link (see Figure~\ref{fig:system}). The transmitter (TX) can sense the underlying time-varying process and  generate a status update at each time slot at a certain energy cost. Status updates are communicated to the receiver (RX) over a time-varying wireless channel. Each transmission attempt of a status update takes constant time, which is assumed to be equal to the duration of one time slot. 

The AoI measures the timeliness of the status information at the receiver, and is defined at any time slot $t$ as the number of time slots elapsed since the generation time $U(t)$ of the most up-to-date packet successfully decoded at the receiver. Formally, the AoI at the receiver at time $t$ is defined as $\Delta^{rx}_t\triangleq \min(t-U(t),\Delta_{max})$, where a maximum value $\Delta_{max}$ on the AoI is imposed to limit the  impact of the AoI on the performance after some level of staleness is reached.

We assume that the channel changes randomly from one time slot to the next in an independent and identically distributed (i.i.d.) fashion, and the instantaneous channel state information is available only at the receiver. We further assume the availability of an error- and delay-free single-bit feedback from the receiver to the transmitter for each transmission attempt. Successful reception of the status update at the end of time slot $t$ is acknowledged by an ACK signal (denoted by $K_t=1$), while a NACK signal is sent in case of a failure (denoted by $K_t=0$).

There are three possible actions $A_t$ the transmitter can take at each time slot $t$: it can either sample and transmit a new status update ($A_t=\new$), remain idle ($A_t=\idle $) or retransmit the last transmitted status update ($A_t= \retx$). If an ACK is received at the transmitter, we can restrict the action space to $\{\idle,\new\}$ as retransmitting an already decoded status update is strictly suboptimal.


We consider the HARQ protocol: that is, the received signals from previous transmission attempts for the same packet are combined for decoding. The probability of error using $r$ retransmissions, denoted by $g(r)<1$, depends on $r$ and the particular HARQ scheme used for combining multiple transmission attempts (an empirical method to estimate $g(r)$ is presented in \cite{harq2003}). As in any reasonable HARQ strategy, we assume that $g(r)$ is non-increasing in the number of retransmissions $r$; that is, $g(r_1) \geq g(r_2)$ for all $r_1 \leq r_2$.  Standard HARQ methods only combine information from a finite maximum number of retransmissions \cite{IEEEstandard}. Accordingly, we consider a truncated retransmission count of a status update, denoted by $R_t$ for the status update transmitted at time $t$, where $R_t \in \{0,\ldots,R_{max}\}$; that is, the receiver can combine information from the last $R_{max}$ retransmissions at most. We also assume that $R_0=0$ so that there is no previously transmitted packet at the transmitter at time $t=0$.


At the end of each time slot $t$, a random amount of energy is harvested and stored in a rechargeable battery at the transmitter, denoted by $E_t \in\mathcal{E} \triangleq \{0,1,\ldots,E_{max}\}$, following  a first-order discrete-time Markov model, characterized by  stationary probabilities $p_E(e_1|e_2)$, defined as $p_E(e_1|e_2)\triangleq Pr(E_{t+1}=e_2|E_t=e_1),~\forall t$. It is also assumed that $p_E(0|e)>0$, $\forall e \in \mathcal{E}$. Harvested energy is first stored in a rechargeable battery with a limited capacity of $B_{max}$ energy units and the energy harvested when the battery is full is lost. The energy consumption for status sensing is denoted by $E^s\in \mathbb{Z}^{+}$, while the energy consumption for a transmission attempt is denoted by $E^{tx} \in \mathbb{Z}^{+}$. 




The battery state at time $t$, denoted by $B_t$, and the energy causality constraints can be written as follows:
\begin{eqnarray}
B_{t+1}=\min(B_t+E_t-(E^s+E^{tx})\mathbbm{1}[A_t=\sense]\nonumber\\-E^{tx}\mathbbm{1}[A_t=\retx],B_{max}), \label{eq:causality1}\\
(E^s+E^{tx})\mathbbm{1}[A_t=\sense]+E^{tx}\mathbbm{1}[A_t=\retx] \leq B_t, \label{eq:causality2}
\end{eqnarray}
where the indicator function $\mathbbm{1}[C]$ is equal to $1$ if event $C$ holds, and zero otherwise. Eqn. \eqref{eq:causality1} implies that the battery overflows if energy is harvested when the battery is full, while Eqn. \eqref{eq:causality2} imposes that the energy consumed by sensing or transmission operations at time slot $t$ is limited by the energy $B_{t}$ available in the battery at the beginning of that time slot. 


The age $\Delta^{tx}_t$ of the most recently generated status update at the transmitter at the beginning of time slot $t$ resets to $1$ if a new status update is generated at time slot $t-1$, and increases up to $\Delta_{max}$ otherwise, i.e., 
\begin{align*}
\Delta^{tx}_{t+1}=
\begin{cases}
1 &\textrm{ if }  A_t=\sense; \\
\min(\Delta^{tx}_t+1,\Delta_{max}) &\textrm{ otherwise. }
\end{cases}
\end{align*}
The AoI of the most recent successfully decoded packet at the receiver at time $t$, $\Delta^{rx}_t$, evolves as follows:
\begin{align*}\Delta^{rx}_{t+1}=
\begin{cases}
\min(\Delta^{rx}_t+1,\Delta_{max}) &\textrm{ if } A_t=\idle \textrm{ or } K_t=0; \\
1 &\textrm{ if } A_t=\sense \textrm{ and } K_t=1; \\
\min(\Delta^{tx}_t+1,\Delta_{max}) &\textrm{ if } A_t=\retx \textrm{ and } K_t=1.
\end{cases}
\end{align*}
We note that $\Delta^{tx}_t$ refers to the number of time slots elapsed since the generation of the most recently sensed status update at the transmitter side, while $\Delta^{rx}_t$ denotes the AoI of the most recently received status update at the receiver side. The system model also implies that whenever a new status update packet is generated, the previous packet at the transmitter is dropped and can not be retransmitted.  The number of retransmissions is zero for a newly sensed and generated status update and increases up to $R_{max}$ as we keep retransmitting the same packet. 
\begin{align*}
R_{t+1}=
\begin{cases}
0 &\textrm{ if }  K_t=1; \\
1 &\textrm{ if } A_t=\sense \textrm{ and } K_t=0; \\
R_t &\textrm{ if }  A_t=\idle; \\
\min(R_t+1,R_{max}) &\textrm{ if } A_t=\retx \textrm{ and } K_t=0. 
\end{cases}
\end{align*}

The state of the system is formed by five components $S_t=(E_{t},B_t,\Delta^{rx}_t,\Delta^{tx}_t,R_t)$. At each time slot, the transmitter knows the state of the system and the goal is to find a policy $\pi$ which minimizes the expected average AoI at the receiver over an infinite time horizon, which is given by:
\begin{align}
J^*\triangleq \min_{\pi}\lim_{T\rightarrow \infty }\frac{1}{T+1}\Exp{\sum_{t=0}^T{\Delta^{rx}_t}} \label{eq:problem}\\
\textrm{subject to } \eqref{eq:causality1} \textrm{ and } \eqref{eq:causality2} \nonumber.
\end{align}

\section{Markov Decision Process (MDP) and RVI}
\label{sec:dynamic}

An average-cost finite-state MDP provides the necessary framework for modeling and solving the AoI minimization problem in \eqref{eq:problem}. An MDP is defined by the quadruple  $\big(\mathcal{S}, \mathcal{A}, $P$, c\big)$ \cite{Puterman_book}: The finite set of states  $(E_t,B_t,\Delta^{rx}_t , \Delta^{tx}_t, R_t)$ is  $\mathcal{S}=\mathcal{E}\times \{0,\ldots,B_{max}\}\times\{1,\ldots,\Delta_{max}\}^2\times \{0,\ldots,R_{max}\}$  and the finite set of actions $\mathcal{A}=\{\idle,\sense,\retx\}$ are already defined. $P$ refers to the transition probabilities, where  $P(s'|s,a) = \Pr(S_{t+1}=s' \mid S_t = s, A_t=a)$ is the probability that action ${\displaystyle a}$  in state ${\displaystyle s}$  at time ${\displaystyle t}$  will lead to state ${\displaystyle s'}$ at time ${\displaystyle t+1}$, which is characterized by the EH statistics and channel error probabilities. The cost function $c: \mathcal{S} \times \mathcal{A} \rightarrow \mathbbm{Z}$, is the AoI at the receiver, and is defined as $c(s,a)=\Delta^{rx}_t$ for any $s\in \mathcal{S}$, $a \in \mathcal{A}$, independent of the action $a$. 

We note that there exists an optimal stationary deterministic policy, $\pi:\mathcal{S}\rightarrow \mathcal{A}$ , for this problem\footnote{For Markov chains corresponding to every stationary policy, there is only one recurrent class as the state $(0,0,\Delta_{max},\Delta_{max},0)$ is reachable from all other states (e.g., every transmission is successful but no EH is harvested for a period of $\max(\Delta_{max},B_{max})$ time slots) from Theorem~8.4.3 of \cite{Puterman_book}.} \cite{Puterman_book}.  In particular, there exists a function $h(s$, called the \textit{differential cost function} for all $s=(e,b,\delta^{rx},\delta^{tx},r) \in \mathcal{S}$, satisfying the following \emph{Bellman optimality equations} for the average-cost finite-state finite-action MDP \cite{Puterman_book}:
\begin{align}
\label{eq:Bellman}
h(s)+J^{*}&=\min_{a\in\{\idle,\sense,\retx\}}\big(\delta^{rx}+\Exp{h(s')|a}\big),
\end{align}
where $s'\triangleq (e',b',{\delta^{rx}}',{\delta^{tx}}',r')$ is the next state obtained from $(e,b,\delta^{rx},\delta^{tx},r)$ after taking action $a$, and $J^*$ represents the optimal achievable average AoI under policy $\pi^*$. Note that the function $h$ satisfying \eqref{eq:Bellman} is unique up to an additive factor, and with selecting this additive factor properly, it also satisfies
\begin{align*}
 h(s)  = \Exp{\sum_{t=0}^\infty (\Delta^{rx}_t - J^{*})\big| S_0=s}  \end{align*}
We also introduce the \textit{state-action cost function}: 
\begin{equation}
Q((e,b,\delta^{rx},\delta^{tx},r),a)\triangleq \delta+\Exp{h(e',b',{\delta^{rx}}',{\delta^{tx}}',r')|a}~.
\label{eq:Bellman2}
\end{equation}
Then an optimal policy, for any $(e,b,\delta^{rx},\delta^{tx},r) \in \mathcal{S}$, takes the action achieving the minimum in \eqref{eq:Bellman2}:
\begin{align}
\label{eq:opt_eta}
\pi^*(e,b,\delta^{rx},\delta^{tx},r) &\in \argmin_{a\in\{\idle,\sense,\retx\}} \big(Q((e,b,\delta^{rx},\delta^{tx},r),a)\big)~. 
\end{align}

An optimal policy solving \eqref{eq:Bellman}, \eqref{eq:Bellman2} and \eqref{eq:opt_eta} defined above can be found by relative value iteration (RVI) for finite-state finite-action average-cost MDPs from Section 8.5.5 of \cite{Puterman_book}:

Starting with an arbitrary initialization of $h_0(s)$, $\forall s \in \mathcal{S}$, and setting an arbitrary but fixed reference state $s^{ref}\triangleq (e^{ref},b^{ref},{\delta^{rx}}^{ref},{\delta^{tx}}^{ref},r^{ref})$,  a single iteration of the RVI algorithm $\forall (s,a) \in \mathcal{S}\times \mathcal{A}$ is given as follows:     \begin{align}
Q_{n+1}(s,a) &\leftarrow \Delta^{rx}_n+	\Exppi{h_n(s')},\\
V_{n+1}(s) &\leftarrow \min_{a}(Q_{n+1}(s,a)),\\
h_{n+1}(s) &\leftarrow {V}_{n+1}(s) -{V}_{n+1}(s^{ref}),
\end{align} 
where $Q_n(s,a)$, $V_n(s)$ and $h_n(s)$ denote the state-action value function, value function and differential value function for iteration $n$, respectively. By Theorem 8.5.7 and Section 8.5.5 of \cite{Puterman_book},  $h_n$ converges to $h$, and $\pi_{n}^*(s) \triangleq \argmin_{a} Q_{n}(s,a)$ converges to $\pi^*(s)$.

\section{A Reinforcement Learning Approach}

In most practical scenarios, channel error probabilities for  retransmissions and the EH characteristics may not be known at the time of deployment, or may change over time. In this section, we assume that the transmitter  does not know the system characteristics \textit{a-priori}, and has to learn them. We employ two different online learning algorithms. First, we employ a value-based RL algorithm, namely GR-learning, which converges to an optimal policy; then, we consider a structured policy search algorithm, namely finite-difference policy gradient, which does not necessarily find the optimal policy but performs very well in practice, as demonstrated through simulations in Section~\ref{sec:simulation}. We also note that GR-learning learns from a single trajectory generated during learning steps while policy gradient uses Monte-Carlo roll-outs for each policy update. Thus, GR-learning is more applicable to real-time systems.

\subsection{GR-Learning with Softmax}
\label{sec:gr}

The literature for average-cost RL is quite limited compared to discounted cost problems  \cite{Sutton1998,Mahadevan1996}.  For the average AoI minimization problem in \eqref{eq:problem}, we employ a modified version of the  \textit{GR-learning} algorithm proposed in \cite{Gosavi2004}, as outlined in Algorithm~1, with \emph{Boltzmann}  (\emph{softmax}) exploration.  The resulting algorithm is called \emph{GR-learning with softmax}.

Notice that, by only knowing $Q(s,a)$, one can find the optimal policy $\pi^*$ using \eqref{eq:opt_eta} without knowing the transition probabilities $P$ characterized by $g(r)$ and $p_E$. Thus, \emph{GR-learning with softmax} starts with an initial estimation of $Q_{0}(s,a)$ and finds the optimal policy by estimating state-action values in a recursive manner. In the $n^{th}$ iteration, after taking action $A_n$, the transmitter observes the next state $S_{n+1}$, and the instantaneous cost value $\Delta^{rx}_n$. Based on this, the estimate of $Q_{n+1}(s,a)$ is updated by a weighted average of the previous estimate $Q_{n}(s,a)$ and the estimated expected value of the current policy in the next state $S_{n+1}$. Moreover, we update the gain $J_n$ at every time slot based on the empirical average of AoI. 

In each time slot, the learning algorithm 
\begin{itemize}
\item observes the current state $S_n \in \mathcal{S}$,
\item selects and performs an action $A_n \in \mathcal{A}$,
\item observes the next state $S_{n+1}\in \mathcal{S}$ and the instantaneous cost $\Delta^{rx}_n$,
\item updates its estimate of $Q(S_n,A_n)$ using the current estimate of $J_{n}$ by
\begin{align}
Q_{n+1}(S_n,A_n)\leftarrow Q_{n}(S_n,A_n) +  \alpha(m(S_n,A_n,n)) \nonumber \\ [\Delta^{rx}_n -J_{n}+Q_{n}(S_{n+1},A_{n+1})-Q_{n}(S_n,A_n)],
\end{align}
where $\alpha(m(S_n,A_n,n))$ is the update parameter (learning rate) in the $n^{th}$ iteration, and depends on the function $m(S_n,A_n,n)$, which is the number of times the state–action pair $(S_n, A_n)$ was visited till the $n^{th}$ iteration.
\item updates its estimate of $J_{n}$ based on the empirical average as follows:
\begin{align}
  J_{n+1}\leftarrow J_n+ \beta(n) \left[\frac{n J_n+\Delta^{rx}_n}{n+1}-J_n\right]  \end{align}
 where $\beta(n)$ is the update parameter in the $n^{th}$ iteration. 
  \end{itemize}

The transmitter action selection method should balance the \textit{exploration} of new actions with the \textit{exploitation} of actions known to perform well. In particular, the \textit{Boltzmann} (\textit{softmax}) action selection method, which chooses each action randomly relative to its expected cost, is used in this paper as follows:
\begin{equation}
    \pi(a|S_n)=\frac{\displaystyle\exp(-Q(S_n,a)/\tau_n)}{\displaystyle\sum_{a'\in\mathcal{A}}{\exp(-Q(S_n,a')/\tau_n)}} \label{eq:softmax}.
\end{equation}
Parameter $\tau$ in \eqref{eq:softmax} is called the temperature parameter and decays exponentially with decay parameter $\gamma$. High $\tau$ corresponds to more uniform action selection (exploration) whereas low $\tau$ is biased toward the best action (exploitation). According to Theorem~2 of \cite{Gosavi2004}, if $\alpha$, $\beta$ satisfy $\sum_{m=1}^{\infty}\alpha(m), \sum_{m=1}^{\infty}\beta(m)\rightarrow \infty$,   $\sum_{m=1}^{\infty}\alpha^2(m), \sum_{m=1}^{\infty}\beta^2(m) <\infty$, $
\lim_{x \to \infty} \frac{\beta(m)}{\alpha(m)}\rightarrow 0$, \textit{GR}-Learning converges to an optimal policy. 

\subsection{Finite-Difference Policy Gradient}
\label{sec:policy_gradient}

GR-learning in Section~\ref{sec:gr} is a value-based RL method, which learns the state-action value function for each state-action pair. In practice, $\Delta_{max}$ can be large, which might slow down the convergence of GR-learning due to a large state-space. 


In this section, we are going to simplify the problem  and obtain a structured possibly sub-optimal policy, which can be learned via the policy gradient method \cite{policy_gradient}.  We make two assumptions on the policy space in order to obtain a more efficient learning algorithm:
\begin{itemize}
    \item We assume that a packet is retransmitted until it is  successfully decoded, provided that there is enough energy in the battery, that is, the transmitter is not allowed to preempt an undecoded packet and transmit a new one. 
    \item The solution to the simplified problem is threshold-type,  that is, 
\begin{align}
A_t= \begin{cases}
\idle &\textrm{ if }  \Delta_t < \mathcal{T}(e,b,\delta^{tx},r) \\
\new  &\textrm{ if }  \Delta_t \geq \mathcal{T}(e,b,\delta^{tx},r) \textrm{ and } r=0 \\
\retx  &\textrm{ if }  \Delta_t \geq \mathcal{T}(e,b,\delta^{tx},r) \textrm{ and } r\neq 0
\end{cases}
\end{align} 
for some $\mathcal{T}(e,b,\delta^{tx},r)$.
\end{itemize}

Note that $A_t=\idle$ if $b<E^{tx}$  ($b<E^{tx}+E^s$) for $r>1$ ($r=1$); that is,  $\mathcal{T}(e,b,\delta^{tx},r)=\Delta_{max}+1$. This ensures that energy causality constraints in \eqref{eq:causality2} hold. Other thresholds will be determined using policy gradient. 

In order to employ the policy gradient method, we approximate the policy by a parameterized smooth function with parameters $\theta(e,b,\delta^{tx},r)$, and convert the discrete policy search problem into estimating the optimal values of some continuous parameters, which can be numerically solved by  stochastic approximation algorithms \cite{Spall2003}.


In particular, with a slight abuse of notation, we let $\pi_{\theta}(e,b,\delta^{rx},\delta^{tx},r)$ denote the probability of taking action $A_t=\sense$ ($A_t=\retx$) if $r=0$ ($r\neq 0$), and  consider the parameterized sigmoid function:
\begin{align}
\pi_{\theta}(e,b,\delta^{rx},\delta^{tx},r)\triangleq \frac{1}{1-e^{-\frac{\delta-\theta(e,b,\delta^{tx},r)}{\tau}}}.
\end{align}
We note that $\pi_{\theta}(e,b,\delta^{rx},\delta^{tx},r) \rightarrow \{0,1\}$ and $\theta(e,b,\delta^{tx},r)\rightarrow \mathcal{T}(e,b,\delta^{tx},r)$ as $\tau \rightarrow 0$. Therefore, in order to converge to a deterministic policy $\pi$, $\tau>0$ can be taken as a sufficiently small constant, or can be decreased gradually to zero.  The total number of parameters to be estimated is $|\mathcal{E}| \times B_{max}\times \Delta_{max}\times R_{max}+1$ minus the parameters corresponding to  $b<E^{tx}$  ($b<E^{tx}+E^s$) for $r>0$ ($r=0$) due to energy causality constraints as stated previously. 

With a slight abuse of notation, we map the parameters $\theta(e,b,\delta^{tx},r)$ to a vector $\overline{\theta}$ of size $d\triangleq |\mathcal{E}| \times B_{max} \times \Delta_{max}\times R_{max}+1$.  Starting with some initial estimates of $\overline{\theta}_0$, the parameters can be updated in each iteration $n$ using the gradients as follows: 
\begin{align}
   \overline{\theta}_{n+1}=\overline{\theta}_n-\gamma(n) ~{\partial J}/{\partial \overline{\theta}_n},
\end{align}
where the step size parameter $\gamma(n)$ is a positive decreasing sequence and satisfies the first two convergence properties given at the end of Section~\ref{sec:gr}. 

Computing the gradient of the average AoI directly is not possible; however, several methods exist in the literature to estimate the gradient \cite{Spall2003}. In particular, we employ the finite-difference policy gradient  \cite{policy_gradient} method. In this method, the gradient is estimated by estimating $J$ at slightly perturbed parameter values. First, a random perturbation vector $D_n$ of size $d$ is generated according to a predefined probability distribution, e.g., each component of $D_n$ is an independent Bernoulli random variable with parameter $q\in (0,1)$. The thresholds are perturbed with a small amount $\sigma>0$ in the directions defined by $D_n$ to obtain $\overline{\theta}_n^{\pm}(e,b,\delta^{tx},r)\triangleq \overline{\theta}_n(e,b,\delta^{tx},r)\pm \sigma D_n$. Then, empirical estimates $\widehat{J}^{\pm}$ of the average AoI corresponding to the perturbed parameters $\overline{\theta}_n^{\pm}$, obtained from Monte-Carlo rollouts, are used to estimate the gradient:
\begin{align}
  {\partial J}/{\partial \overline{\theta}_n} \approx (D_n^{\intercal} D_n)^{-1} D_n^{\intercal} \frac{(\widehat{J}^+-\widehat{J}^-)}{2 \sigma}. 
\end{align}
where $D_n^{\intercal}$ denotes the transpose of vector $D_n$.

\section{Simulation Results}
\label{sec:simulation}

In this section, we provide numerical results for all the proposed algorithms, and compare the achieved average AoI. Motivated by previous research on HARQ \cite{hybrid2001}, \cite{harq2003}, \cite{IEEEstandard}, we assume that the decoding error reduces exponentially with the number of retransmissions, that is, $g(r)\triangleq p_0 \lambda^{r}$ for some $\lambda \in (0,1)$, where $p_0$ denotes the error probability of the first transmission and $r$ is the retransmission count (set to $0$ for the first transmission). 
The exact value of the rate $\lambda$ depends on the particular HARQ protocol and the channel model. Following the \emph{IEEE 802.16} standard\cite{IEEEstandard}, the maximum number of retransmissions used for decoding is set to $R_{max}=3$. In the following experiments, $\lambda$ and $p_0$ are set to $0.5$. $E^{tx}$ and $E^{s}$ are both assumed to be constant and equal to 1 unit of energy unless otherwise stated. $\Delta_{max}$ is set to $40$. 

We choose the exact step sizes for the learning algorithms by fine-tuning in order to balance the algorithm stability in the early time steps with nonnegligible step sizes in the later time steps.  In particular, we use step size parameters of  $\alpha(m),\beta(m),\gamma(m)=y/(m+1)^z$,  where $0.5<z\leq 1$ and $y>0$ (which satisfy the convergence conditions) and choose $y$ and $z$ such that the oscillations are low and the convergence rate is high. We have observed that a particular choice of parameters gives similar performance results for scenarios addressed in simulations results. 

\subsection{Uncorrelated EH}

\begin{figure}
\centering
\includegraphics[scale=0.35]{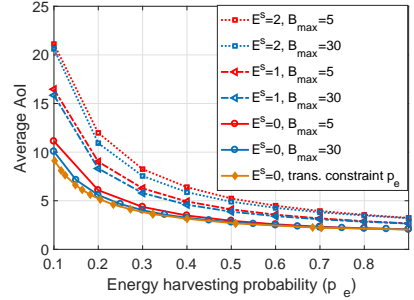}
\caption{Average AoI for different $B_{max}$, $E^s$ and $p_e$ values when EH is i.i.d. and $E^{tx}=1$.}
\label{fig:effectofB}
\end{figure}

We first investigate the average AoI with HARQ when the EH process, $E_t\in \mathcal{E}=\{0,1\}$, is i.i.d. over time with probability distribution  $Pr(E_t=1)=p_e$, $\forall t$. The RVI algorithm in Section~\ref{sec:dynamic} is employed, and the effects of the battery capacity $B_{max}$, energy consumption of sensing $E^s$, and $p_e$ on the average AoI are shown in Figure~\ref{fig:effectofB}. As expected, the average AoI increases with decreasing $B_{max}$, decreasing $p_e$ and increasing $E^s$.  We note that, when $E^s=0$ and $B_{max}=\infty$, the problem defined in~\eqref{eq:problem} corresponds to minimizing the average AoI under an average transmission rate constraint $p_e$, studied in \cite{wcnc_paper,journal}. The average AoI under average transmission rate constraint ($B_{max}=\infty$) is also shown in Figure~\ref{fig:effectofB}.

\begin{figure}
    \centering
    \includegraphics[scale=0.35]{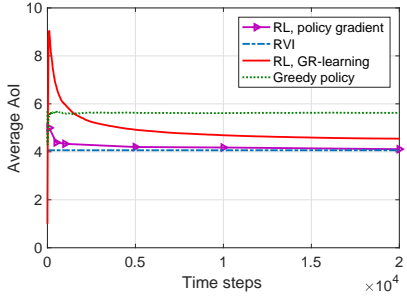}
    \caption{Performance of RL algorithms when $B_{max}=5$, $E^s,E^{tx}=1$, and $p_e=0.5$.}
    \label{fig:iid05}
\end{figure}

Figure~\ref{fig:iid05} shows the evolution of the average AoI over time when the average-cost RL algorithms are employed. As a baseline, we have also included the performance of a greedy policy, which sends a new status update whenever there is sufficient energy for both sensing and transmission. It
retransmits the last transmitted status update
when the energy in the battery is sufficient
only for transmission, and it remains idle otherwise; that is, $A_t=\sense$ if $B_t\geq E^{tx}+E^{s}$,  $A_t=\retx$ if $E^{tx}\leq B_t< E^{tx}+E^{s}$ and $A_t=\idle$ if $B_t<E^{tx}$. It can  be observed that the average AoI achieved by the proposed RL algorithms, converge to values close to the one obtained from the RVI algorithm, which has \emph{a priori} knowledge of $g(r)$ and $p_e$, while the AoI of the greedy algorithm is significantly higher. Although the policy gradient algorithm based on threshold policy does not allow preemption of an undecoded status update, it performs better than GR-learning since it tries to learn significantly smaller number of threshold values (i.e., $\Delta_{max}\times  B_{max}\times R_{max}+1$) than GR-learning which learns one value for each state-action pair (i.e., $\Delta_{max}^2 \times B_{max}\times (R_{max}+1) \times |\mathcal{A}| $).  



\subsection{Temporally Correlated EH}

Next, we investigate the performance when the EH process has temporal correlations. A symmetric two-state Markovian EH process is assumed, such that $\mathcal{E}=\{0,1\}$  and $Pr(E_{t+1}=1|E_t=0)=Pr(E_{t+1}=0|E_t=1)=0.3$. That is, if the transmitter is in harvesting state, it is more likely to continue harvesting energy, and vice versa for the non-harvesting state. 

\begin{figure}
\centering
\begin{subfigure}{0.5\textwidth}
\centering
\includegraphics[scale=0.35]{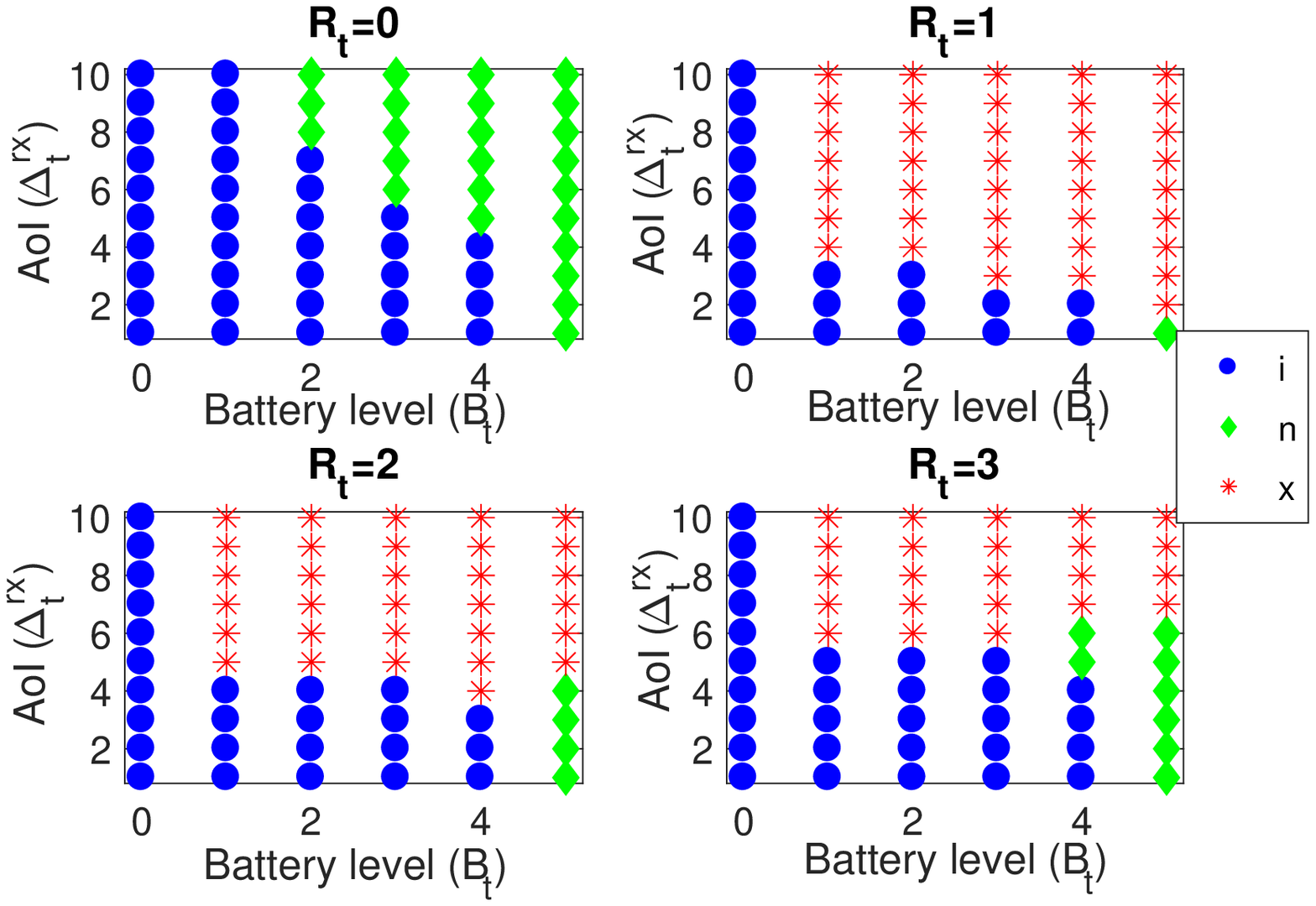}
\caption{$E_t=1$ }
\end{subfigure}
\begin{subfigure}{0.5\textwidth}
\centering
\includegraphics[scale=0.35]{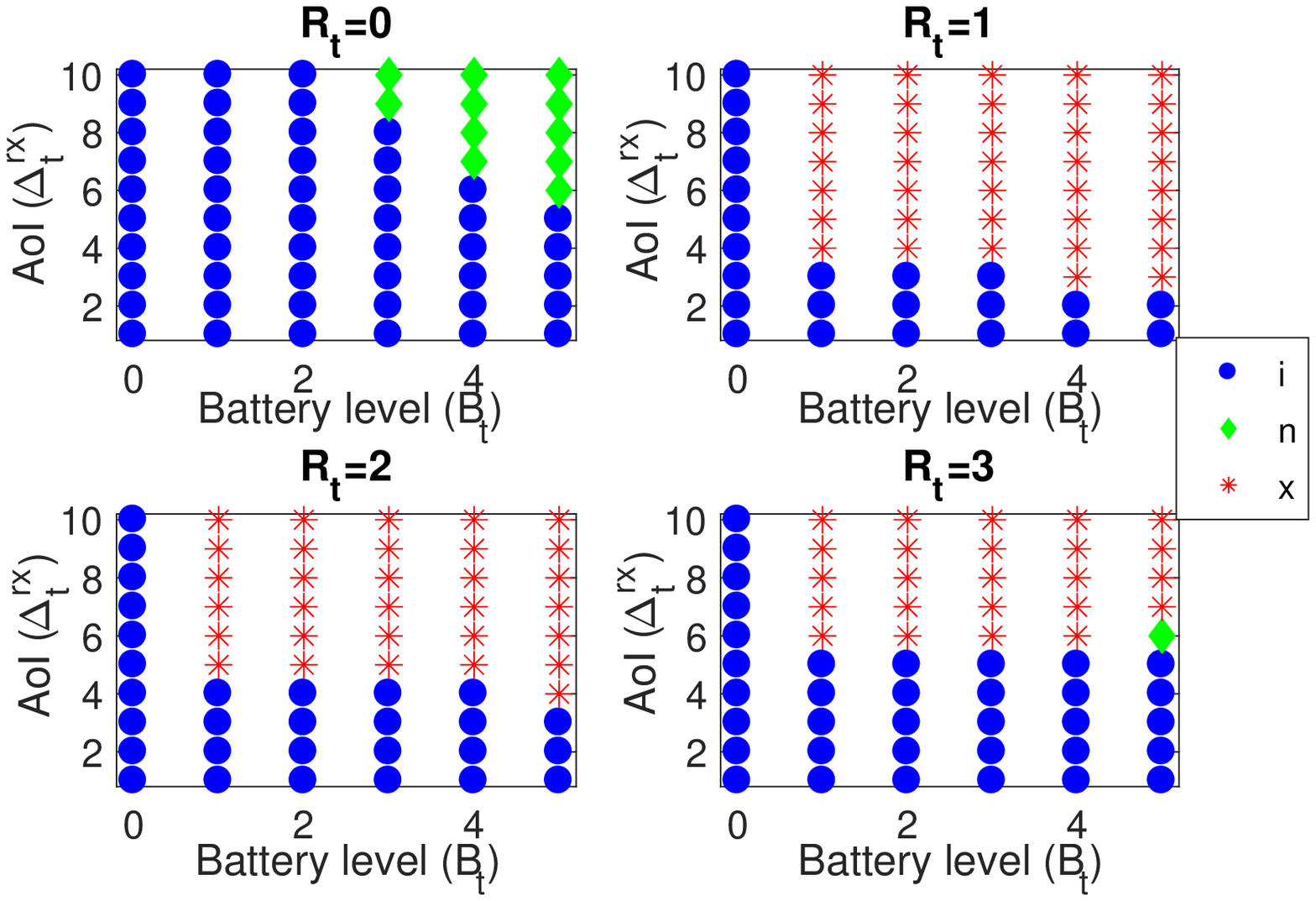}
\caption{$E_t=0$ }
\end{subfigure}
\caption{Optimal policy for $B_{max}=5$, $R_{max}=3$, $p_E(1,1)$, $p_E(0,0)=0.7$, $E^s,E^{tx}=1$ and $\Delta^{tx}_t=R_t+1$. The decoding error probabilities are given by $g(r)= 2^{-(r+1)}$. }
\label{fig:policy}
\end{figure}

Figure~\ref{fig:policy} illustrates the policy obtained by RVI. As it can be seen from the figure, the resulting policy is less likely to transmit if the battery level or the AoI is low. Moreover, the policy tends to retransmit the previous update rather than sensing a new update when the battery level is low and the AoI is high. When the system is in the non-harvesting state (i.e., $E_t=0$), the transmitter is more conservative in transmitting the status updates compared to the case $E_t=1$, e.g., it might not transmit even if the battery is full depending on the AoI level.  

\begin{figure}
    \centering
    \includegraphics[scale=0.35]{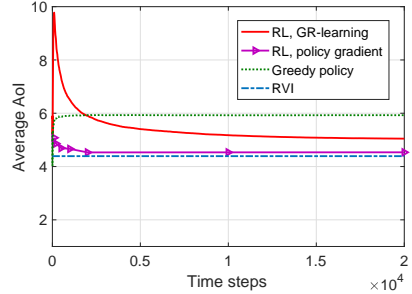}
    \caption{The performance of RL algorithms when $B_{max}=5$, $p_E(1,1)$, $p_E(0,0)=0.7$ and $E^s,E^{tx}=1$.}
    \label{fig:learn_corr_time}
\end{figure}

Figure~\ref{fig:learn_corr_time} shows the evolution of the average AoI over time when the average-cost RL algorithms are employed. It can  be observed again that the average AoI achieved by the learned threshold parameters in Section~\ref{sec:policy_gradient}, denoted by policy gradient in the figure, performs very close to the one obtained from the RVI algorithm, which has \emph{a priori} knowledge of $g(r)$ and $p_e$. GR-learning, on the other hand, outperforms the greedy policy but converges to the optimal policy much more slowly, and the gap between the two RL algorithms is even longer compared to the i.i.d. case. Tabular methods in RL, like GR-learning, need to visit each state-action pair infinitely often for RL to converge \cite{Sutton1998}.  GR-learning in the case of  temporally correlated EH does not perform as well as in the i.i.d. case since the state space becomes larger with the addition of the EH state.   

\begin{figure}
    \centering
    \includegraphics[scale=0.35]{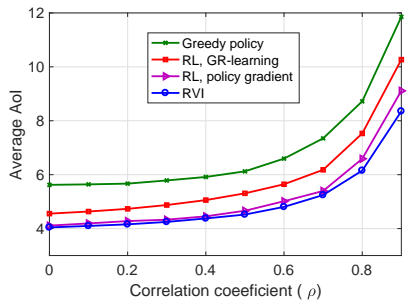}
    \caption{The performance of RL algorithms obtained after $2\cdot10^4$ time steps and averaged over $1000$ runs for different temporal correlation coefficients.}
    \label{fig:learn_corr}
\end{figure}

Next, we investigate the impact of the burstiness of the EH process, measured
by the correlation coefficient between $E_t$ and
$E_{t+1}$. Figure~\ref{fig:learn_corr} illustrates the performance of the proposed RL algorithms for different correlation coefficients, which can be computed easily for the 2-state symmetric Markov chain; that is, $\rho\triangleq (2p_E(1,1)-1)$. Note that $\rho=0$ corresponds to the i.i.d. EH with $p_e=1/2$.   We note that the average AoI is minimized by transmitting new packets successfully at regular intervals, which has been well investigated in previous works \cite{Tan2015,wcnc_paper,Yates2015}. Intuitively, for highly correlated EH, there are either successive transmissions or successive idle time slots, which increases the average AoI. Hence, the AoI is higher for higher values of $\rho$.  Figure~\ref{fig:learn_corr} also shows that both RL algorithms result in much lower average AoI than the greedy policy and 
policy gradient RL outperforms GR-learning since it benefits from the structural characteristics of a threshold policy. 

\section{Conclusions}\label{sec:conclusion}
We have considered an EH system with a finite size battery and investigated scheduling policies transmitting time-sensitive data over a noisy channel with the average AoI as the performance measure, which quantifies the timeliness of the data available at the receiver.  In addition to identifying a RVI solution for the optimal policy when the system characteristics are known, efficient RL algorithms are also presented for practical applications when the system characteristics may not be known in advance. The effects of battery size, EH characteristics and the HARQ structure on the average AoI are investigated through numerical simulations. The algorithms adopted in this paper are relevant to other systems concerning the timeliness of information or those powered by renewable energy sources. 

\bibliography{ageof8}




\end{document}